# 3D dose prediction for Gamma Knife radiosurgery using deep learning and data modification

Binghao Zhang[1], Aaron Babier[1], Timothy C.Y. Chan[1], Mark Ruschin[2]

[1] Department of Mechanical and Industrial Engineering, University of Toronto, Toronto, Canada
[2] Department of Radiation Oncology, Sunnybrook Health Sciences Centre, University of Toronto, Toronto, Canada

E-mail: binghao.zhang@mail.utoronto.ca

**Abstract**

*Purpose:* To develop a machine learning-based, 3D dose prediction methodology for Gamma Knife (GK) radiosurgery. The methodology accounts for cases involving targets of any number, size, and shape.
*Methods:* Data from 322 GK treatment plans was modified by isolating and cropping the contoured MRI and clinical dose distributions based on tumor location, then scaling the resulting tumor spaces to a standard size. An accompanying 3D tensor was created for each instance to account for tumor size. The modified dataset for 272 patients was used to train both a generative adversarial network (GAN-GK) and a 3D U-Net model (U-Net-GK). Unmodified data was used to train equivalent baseline models. All models were used to predict the dose distribution of 50 out-of-sample patients. Prediction accuracy was evaluated using gamma, with criteria of 4%/2mm, 3%/3mm, 3%/1mm and 1%/1mm. Prediction quality was assessed using coverage, selectivity, and conformity indices.
*Results:* The predictions resulting from GAN-GK and U-Net-GK were similar to their clinical counterparts, with average gamma (4%/2mm) passing rates of 84.9 ± 15.3% and 83.1 ± 17.2%, respectively. In contrast, the gamma passing rate of baseline models were significantly worse than their respective GK-specific models (p < 0.001) at all criterion levels. The quality of GK-specific predictions was also similar to that of clinical plans.
*Conclusion:* Deep learning models can use GK-specific data modification to predict 3D dose distributions for GKRS plans with a large range in size, shape, or number of targets. Standard deep learning models applied to unmodified GK data generated poorer predictions.

# 1. Introduction

Gamma Knife (GK) radiosurgery (GKRS) is a form of radiotherapy that precisely treats abnormalities within the brain using narrow beams of radiation. GKRS is an effective treatment for a wide array of diseases including benign tumors, malignant tumors, vascular abnormalities, and functional disorders [1]. Conventional processes to generate GKRS treatment plans are time-consuming for clinicians, which has motivated several studies to explore new approaches like inverse planning [2,3]. However, a major limitation of inverse planning is that it requires human intervention to tune parameters and personalize the resulting treatment plans.

There exist automated planning methods for other modalities that can generate patient specific parameters for inverse planning [4,5]. An integral part of these approaches is a machine learning (ML) method that produces dose predictions using patient images. There is also a small set of models that incorporate additional patient features (e.g., age, histology) to account for patient outcomes [4,5]. In general, automated planning approaches that use predicted dose distributions are called knowledge-based planning (KBP) pipelines. A KBP pipeline is typically presented as a two-stage process that leverages information from previous treatment plans to produce high-quality treatment plans for new patients without human intervention. The first stage is a dose prediction model that learns the relationship between dose and delineated medical images from previous plans. The second stage is an optimization model that generates a treatment plan from the predicted dose distribution.

Many recent advances in KBP have focused on 3D dose prediction using neural networks [4,5]. These approaches have primarily been developed and tested for intensity-modulated radiotherapy (IMRT) and volumetric modulated arc therapy (VMAT) [6-9]. However, GKRS presents three unique challenges that necessitate a new approach for dose prediction. First, there is a large range in treatment target size. Many large targets (e.g., post-operative metastases or benign tumors) are up to 25 times the diameter of small targets (e.g., small intact brain metastases) [10]. This variation in target size requires a prediction model that can adequately accommodate both the smallest and largest targets. Second, GKRS cases can have a relatively large number of targets (e.g., more than 30) with multiple dose prescription levels. As a result, the impact of dose to one target on another can vary drastically between patients. Third, targets are often separated by large amounts of healthy brain tissue. A standard ML approach that considers the whole treatment volume would require a low spatial resolution (i.e., large voxel volumes) to accommodate computational memory limits associated with large neural networks, which would be inadequate for GKRS because it must be planned with a high spatial resolution (i.e., small voxel volumes). These factors further increase both the complexity and spatial resolution requirements of the model.

In this paper, we develop a novel GKRS dose prediction approach. This is an important first step towards creating an automated GKRS planning pipeline since the quality of plans produced by such a pipeline is positively correlated with the quality of the dose predictions [11]. Our approach accommodates any size, number, and shape of targets without compromising the spatial resolution of the predicted dose. The proposed approach involves a novel GKRS-specific data modification method, an upscaling step, and construction of a distance tensor to relate each target back to its size. We demonstrate accuracy on a series of historically treated patient cases. Our high-quality predictions could be used to estimate parameters for inverse optimization models that generate high-quality treatment plans [6].

# 2. Methods

Our methods consisted of five main steps: (2.1) extracting clinical treatment plan data, (2.2) modifying plan image data, (2.3) tailoring existing neural network models for GKRS, (2.4) training dose prediction models, and (2.5) evaluating model dose predictions.

*2.1 Data Extraction*

This research ethics board approved study involved retrospective access to radiotherapy plans for 322 patients who were treated at Sunnybrook Health Sciences Centre. From each plan, we extracted the MRI images, 3D dose distributions, and target contours. All target contours were delineated for treatment by a radiation oncologist on high-resolution MRIs. To visualize the heterogeneity of our dataset, we plotted the distribution of the target size, number of isocentres, number of targets, and prescription dose in a histogram.

*2.2 Data Processing*

The data was processed for our GKRS dose prediction in four major ways, which are summarized in **Figure 1** and explained in the remainder of this section. Patient data was first processed into a format that was amenable for computer vision models (e.g., consistent nomenclature, align data on a voxel grid). Most notably we converted each target contour into a mask that labelled voxels in healthy tissue with 0 and voxels in targets with its prescription dose (e.g., 25 Gy). These masks enabled our dose prediction models to handle plans with a wide range of dose prescription levels that are common in GKRS. This standard pre-processing was applied to all our data and the resulting dataset was used to train and test our baseline models. We developed three additional pre-processing techniques for our GRKS specific approach.

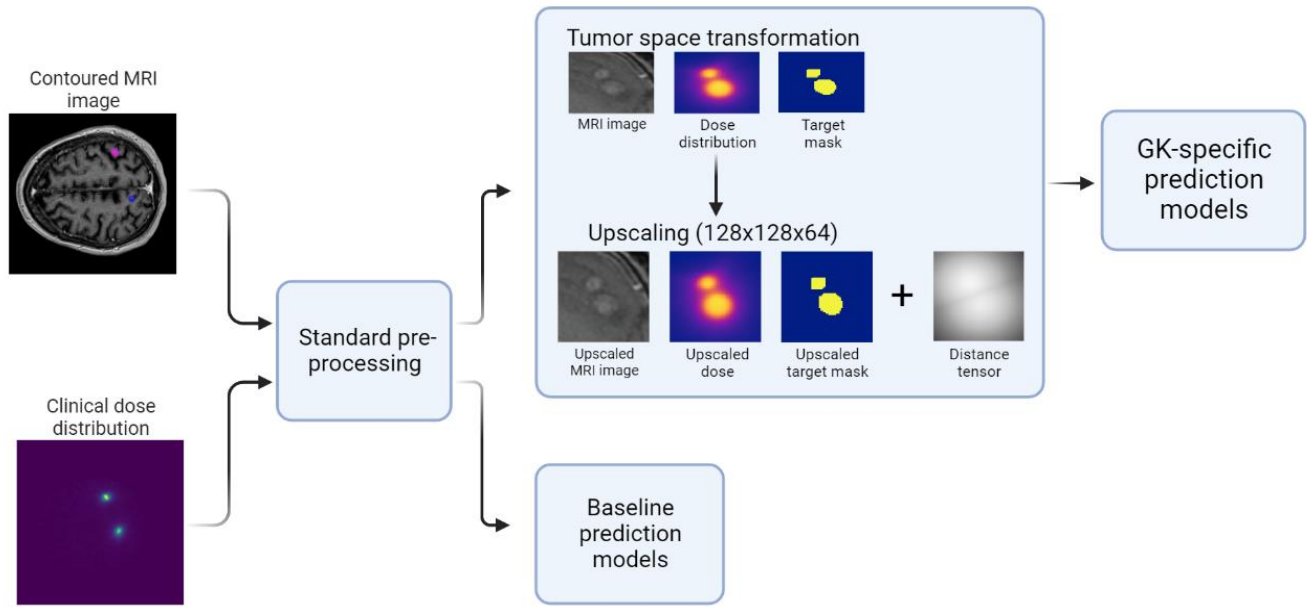

**Figure 1**: An overview of our workflow and the data modification techniques used in this study. Our GK-specific data modification includes transforming patient data with a novel tumor space transformation and an upscaling method. Then we create a new feature that we call a distance tensor to quantify the distance between tissue and targets.

First, we developed *tumor spaces*, which were engineered to isolate small volumes surrounding targets. Specifically, the tumor spaces were the smallest bounding box that contained at least one target surrounded by 1 cm of padding. To ensure that the dosimetric interactions between close targets were captured, any targets within 1 cm of each other were taken together in one tumor space, which is shown by the example in **Figure 1**. We sampled these tumor spaces from the MRI, dose distribution, and target masks of each case to create a training set of 628 tumor spaces from 272 plans. Similarly, we created a testing set of 129 tumor spaces from the 50 plans in the test set.

Second, we developed an upscaling technique to ensure consistent dimensionality across tumor spaces. Inconsistent dimensions normally present a challenge for computer vision models because the models are initialized to expect data with predefined dimensions. To accommodate the range of tumor space dimensions, all data was upscaled using spline interpolation to fit into a 128 x 128 x 64 voxel tensor. A 128 x 128 x 64 tensor size was chosen to balance image detail and training time. The final upscaled tensors included the cropped MRI images, dose distributions, and target masks within each respective tumor space.

Third, for each tumor space we engineered *distance tensors*, which were designed to account for the distance between each voxel and its nearest target. Each element in the distance tensor represented a voxel and had a value equal to the Euclidean distance $d$ between that voxel $v$ and its nearest target centroid $t$. The measure was calculated with respect to all the target centroids $t \in T$ within the patient. It was evaluated over all three spatial dimensions, indexed by $i$. Specifically, the value of each element in the distance tensor was calculated as

$$d = \min_{t \in T} \sqrt{\sum_{i=1}^{3}(v_i - t_i)^2} \ .$$

*2.3 Model Architectures*

Our approach builds on the success of existing neural network models from the IMRT and VMAT literature [6,7,12,13]. Specifically, we adapted the architectures used in previous dose prediction approaches to fit the data size and structure of GKRS. Full details of the model architecture are presented in the accompanied supplement. We implemented two types of models in this study, a U-Net and a generative adversarial network (GAN). The U-Net used a standard 3D architecture to generate a 3D dose using contoured MRI images [14]. A mean squared error loss function was used to train the U-Net. The GAN used a pix2pix architecture [14] to combine the same architecture as our U-Net model with a discriminator, which is a second neural network within the GAN that predicted the likelihood that a dose distribution was from a clinical plan or generated by the U-Net. Both neural networks within the GAN were trained simultaneously such that predictions from the discriminator were used to improve the dose produced by the U-Net model within the GAN via a typical GAN loss function. A binary cross entropy loss function was used for the discriminator model.

*2.4 Model Training and Prediction*

The modified MRI images, target masks, 3D dose distributions, and distance tensors were used to train two GKRS specific dose prediction models, one with a GAN architecture (GAN-GK) and another with only a 3D U-Net architecture (U-Net-GK). To accommodate different prescription doses between cases, clinical dose distributions were normalized relative to its nominal prescription dose prior to training. Baseline models for GAN (GAN-Baseline) and 3D U-Net (U-Net-Baseline) were trained on patient data without GRKS specific processing. The networks were developed in Python 3.7 using TensorFlow 1.12.3.

All models were trained using the same 272 plans in our training dataset. Each model was also trained for 200 epochs on a Nvidia 1080 Ti GPU with 12 GB of memory, which took approximately 6.5 and 3 days for the GAN and U-Net models, respectively. Additionally, all optimization was done via gradient descent with using the Adam optimizer with momentum parameters $\beta_1 = 0.5$, $\beta_2 = 0.999$, and a learning rate of 0.0002. These hyperparameters were selected because they have been effective for a variety of other applications and additional tuning was computationally expensive [14]. The model was trained with a batch size of eight, which was the largest size we could use due to computational limitations.

Predicted 3D dose distributions for the 50 test plans were generated using each model. Dose predictions generated by GAN-GK and U-Net-GK were scaled back to their original target size and prescription dose, and the predictions for all tumor spaces in the patient were combined to recreate a full 3D dose distribution. A dose of zero was assigned to all voxels that were excluded from all tumor spaces, and the average dose was used for voxels with overlapping tumor spaces.

*2.5 Analysis*

To evaluate the accuracy of the dose distribution predictions relative to the clinical delivered dose, a global 3D gamma analysis was used [15,16]. For this analysis, we used four *agreement criteria* that have been used in other GKRS evaluations (4%/2 mm, 3%/3 mm, 3%/1 mm, and 1%/1 mm) [17-19]. A l*ow-dose threshold* equal to 5% of the maximum dose was used to compute the gamma passing rate for each patient. A two-tailed Wilcoxon signed-rank test was used to compare the gamma passing rate of the predictions made with and without data modification, with $p < 0.05$ being considered significant.

Further analysis using a 4%/2 mm gamma passing rate was done to explore where the GKRS specific predictions were most successful and to identify where future improvements are needed. For the purposes of this analysis, each target was divided into three regions: i) the inside, which included all the voxels in the target mask; ii) the periphery, which included all voxels within a two-voxel ring around each target; and iii) the outside, which included the remaining voxels in the tumor space.

To evaluate prediction quality, the coverage, selectivity, and conformity indices [20] were calculated for each target and compared to the same indices for the clinical doses. To compare the difference in quality between GKRS specific predictions and their baseline counterparts, the absolute conformity index difference between predicted and clinical plans was calculated and compared using a two-tailed Wilcoxon signed-rank test, with a significance level of 0.05.

**3. Results**

*3.1 Summary of Clinical Plan Data*

**Figure 2** summarizes the dataset that was used to train and test the models. There was a large range in the size of the targets, number of isocenters per target, and prescription dose. The number of targets per patient ranged from 1 to 26, and the types of targets included brain metastases (treated in 1 to 5 fractions) and acoustic neuromas (treated in 1 fraction). There was a large range in target volumes (34 to 184750 voxels, 0.0085 cc to 46.1875 cc), number of isocenters (1 to 57), and target dose prescriptions (4 to 27.5 Gy). Over 37% and 5% of all targets also had diameters exceeding 2 cm and 4 cm, respectively.

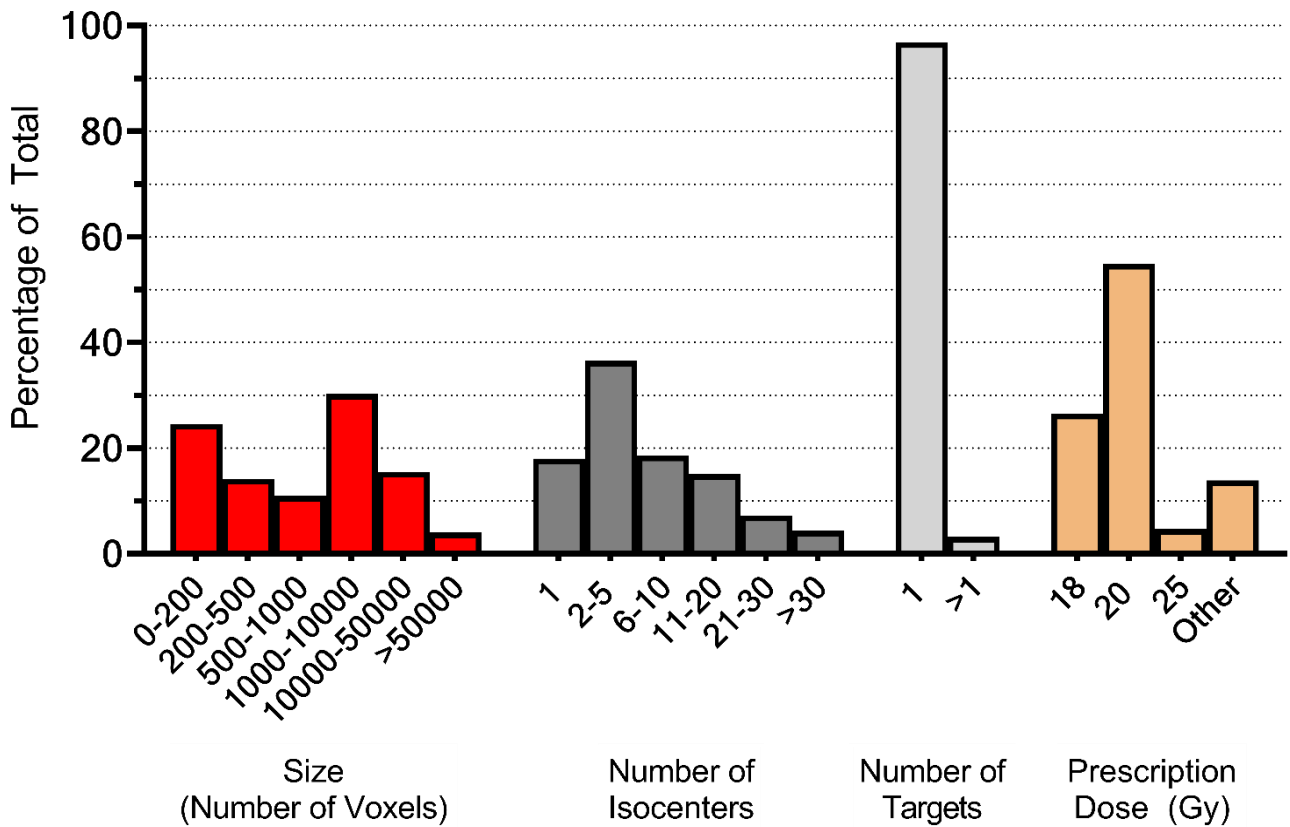

**Figure 2**: Characteristics of the dataset used to train and test the model.

*3.2 Accuracy of Predicted GK-specific 3D Dose Distributions*

**Figure 3** shows the distribution of the gamma passing rate of the predictions for various levels of gamma criteria with respect to the clinical dose. Across all criteria levels, both the GAN-GK and U-Net-GK achieved gamma passing rates that were significantly higher (i.e., better) than that of the GAN-Baseline (Z = -7.37, p < 0.001) and U-Net-Baseline (Z = -7.33, p < 0.001). This result indicates that the GKRS specific approaches produce dose that is more similar to clinical dose than standard baseline approaches. We also found that the performance of each GKRS-specific approach was comparable. For example, compared to the clinical dose using the 4%/2mm gamma criterion, the GAN-GK and U-Net-GK achieved average gamma passing rate of 84.9 ± 15.3% and 83.1 ± 17.2%, respectively; with a 1%/1mm gamma criterion, which is much stricter than the 4%/2mm criterion, GAN-GK and U-Net-GK both achieved much lower average passing rates of 25.2 ± 11.6% and 24.4 ± 11.3%, respectively.

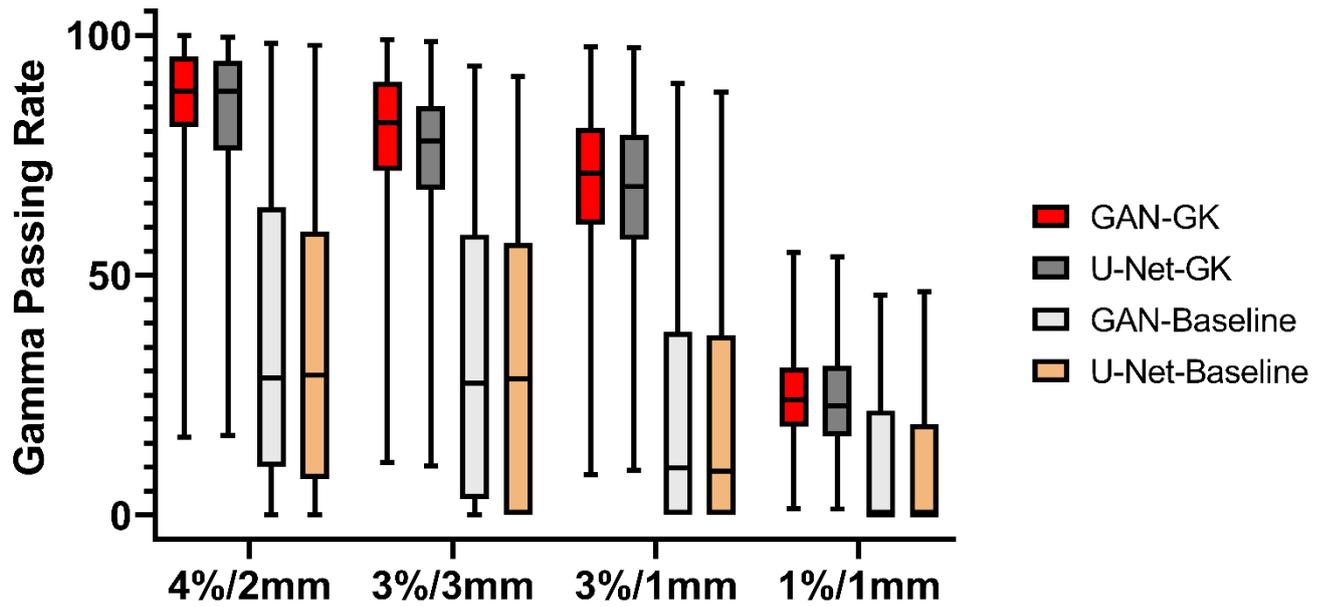

**Figure 3:** The distribution of gamma passing rates for all models at four gamma criterion levels.

With regards to the GKRS specific predictions, the sub-analysis of gamma passing rate of both models showed that the inside of target performed slightly better than the periphery on average, with 82.2 ± 19.5% of the voxels passing compared to 79.8 ± 16.4%. The voxels outside of the target performed the best, with an average passing rate of 91.6 ± 10.7%.

*3.3 Quality of Predicted GK-specific 3D Dose Distributions*

**Table 1** shows the mean and standard deviation for the coverage index, selectivity index, conformity index, and absolute conformity difference for the predictions with respect to the clinical dose. Overall, the GKRS specific approach dominated their baseline alternatives in terms of the coverage, selectivity, and conformity indices. Both the GAN-GK and U-Net-GK predicted doses with coverage, selectivity, and conformity indices that were within 8% of the clinical doses. This result implies that the predictions were very similar to the clinical doses in quality, with an average absolute conformity difference of 0.086 ± 0.11 and 0.092 ± 0.11 for GAN-GK and U-Net-GK, respectively. In contrast, the average conformity of baseline predictions was significantly worse than their corresponding clinical plans, with an average absolute conformity difference of 0.177 ± 0.16 and 0.189 ± 0.17 for GAN-Baseline and U-Net-Baseline, respectively.

|  | **Clinical** | **GAN-GK** | **U-Net-GK** | **GAN-Baseline** | **U-Net-Baseline** |
|---|---|---|---|---|---|
| **Coverage index** | 0.979 ± 0.02 | 0.952 ± 0.11 | 0.968 ± 0.12 | 0.863 ± 0.21 | 0.861 ± 0.22 |
| **Selectivity index** | 0.554 ± 0.22 | 0.597 ± 0.22 | 0.539 ± 0.21 | 0.527 ± 0.21 | 0.542 ± 0.18 |
| **Conformity index** | 0.546 ± 0.22 | 0.560 ± 0.20 | 0.513 ± 0.20 | 0.452 ± 0.22 | 0.474 ± 0.23 |
| **Absolute conformity index difference** | N/A | 0.086 ± 0.11 | 0.092 ± 0.11 | 0.177 ± 0.16 | 0.189 ± 0.17 |

**Table 1:** Average and standard deviation in coverage index, selectivity index, conformity index, and absolute conformity index difference (compared to clinical) for the 3D dose predictions of 50 out-of-sample patients.

*3.4 Visual Comparison of GK-specific Predictions to Baseline Predictions*

**Figure 4** shows an example of predictions made using GK-specific models compared to predictions made using baseline models. The example shows two sample patients (one in each row) to showcase the model performance in different situations. The example highlights the impact of the data modification pipeline, which enables high resolution dose predictions. In addition, predictions made using the baseline models often resulted in predictions with unrealistically low dose to small targets, as seen in **Figure 4f**.

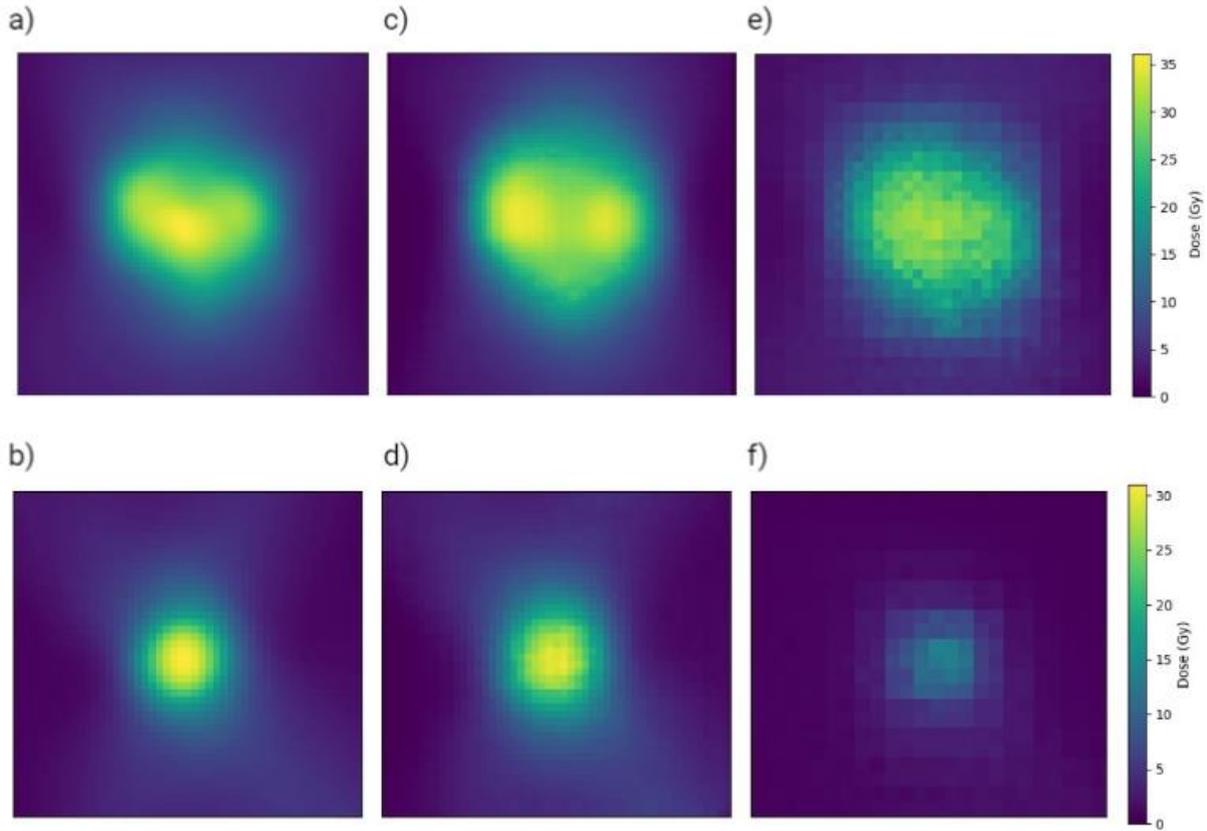

**Figure 4:** a-b) Clinical dose distributions. c) U-Net-GK dose prediction. d) GAN-GK dose prediction. e) U-Net-Baseline dose prediction f) GAN-Baseline dose prediction. As can be seen, predictions made using baseline models are of much lower resolution and sometimes result in low- or no-dose predictions.

## 4. Discussion

In this study, we present novel data modification techniques to facilitate 3D dose prediction for GKRS. We demonstrated that separating the prediction of a full dose distribution into several smaller predictions enables deep learning models to produce more accurate and reliable predictions than those obtained from off-the-shelf methods. Of note, our novel methodology was effective on a heterogenous patient population with a large range of target shapes and sizes. This approach serves as a necessary first step towards developing an KBP pipeline for GKRS that can be adapted for use in any GKRS clinic.

Using the modified data, predictions from GAN-GK and U-Net-GK achieved gamma passing rates similar to or better than those achieved by comparable models in other disease sites [6-8]. For example, a recent study that developed approaches to predict 3D dose distributions of rectal cancer IMRT plans achieved gamma passing rates between 81-90% with a gamma criterion of 3%/5mm [7], which is comparable to our GK-specific approaches that achieved gamma passing rates of 83-85% with a gamma criterion of 4%/2mm. The similarity of the predictions arising from GAN-GK and U-Net-GK to their clinical counterparts is encouraging given the ranges in target size, shape, and quantity among the GKRS plans in our dataset.

While the prediction performs well with looser criteria, when distance-to-agreement and dose difference are restricted to 1%/1mm the predictions are relatively poor with average gamma passing rates of $25.2 \pm 11.6\%$ and $24.4 \pm 11.3\%$ for GAN-GK and U-Net-GK, respectively. However, it seems that the primary factor for this fall in passing rate is due to the stricter dose difference criteria. When the distance-to-agreement criteria is lowered from 3mm to 1mm, with a dose difference of 3%, the passing rate only experienced an average of 10.3% and 8.7% drop for GAN-GK and U-Net-GK, respectively. These results indicate that the methodology can produce predictions which are similar in shape to their clinical counterparts. This is good for GKRS where spatial resolution has relatively high clinical relevance due to steep dose gradients and small targets. In contrast, while predictions appear less likely to match the intensity on a voxel-by-voxel basis – likely due to the small voxel volumes coupled with steep dose gradients – achieving a more accurate dose-agreement is less clinically important because dose is often prescribed to an isodose line in the 50-60% range.

We included several gamma criteria to compliment similar studies in the GKRS literature that compare the similarity of new dose distributions to their clinical counterparts. Our gamma analysis quantified the dosimetric accuracy of predictions in terms of different spatial resolution by varying the spatial portion of the gamma criteria between 1mm and 3mm and the dose portion between 1% and 4%. Across all gamma criteria, the predictions made using GAN-GK and U-Net-GK perform significantly better than baseline predictions. The lower standard deviation on the gamma passing rates of GAN-GK and U-Net-GK predictions also indicate greater consistency. Since better dose predictions are more likely to lead to higher quality plans [11], the presented prediction methodology would serve well as the first stage of a two-stage GKRS KBP pipeline.

Our novel approach for dose prediction is centred around GKRS-specific data modification. This focus is different from many previous studies that focus on developing new architectures [6,7,9,12,13]. As the contributions are focused on the data modification process, we did not fully explore other factors that can improve the predictions such as hyperparameters tuning, tensor sizes, and training duration. The results of this study demonstrate that existing dose prediction models can be tailored for GKRS by data modification alone. This enables us to leverage approaches from the rich dose prediction literature that covers other sites and modalities [6,7,13,21-23]. Most of those studies used a GAN or U-Net architecture. While our GAN model (i.e., GAN-GK) produced marginally better predictions than the U-Net model (i.e., U-Net-GK), a result similar to previous studies [13], it also required more than double the training time of the U-Net model (6.5 days versus 3). As such, training and cross-validation of a U-Net model is more practical for future GKRS datasets.

There are several benefits to leveraging data modification techniques in the training process. First, the training data can use all the pixels stored in the native treatment image without exceeding computational memory constraints. This facilitates models that generate high-resolution dose predictions, as seen in **Figure 3**. Second, using tumor spaces generates more unique data points for the training set. In our case, tumor spaces transformed our training dataset of 272 plans into a set of 628 tumor spaces that were used to train our GK-specific models. We conjecture that increasing the number of data points in the training set enabled the models to generalize better with higher-quality predictions. Lastly, data modification provides flexibility for the shape of plan image data. Specifically, our approach eschews the need for consistent dimensions because we crop and resize the data to consistent dimensions using interpolation, which makes the approach adaptable to variations in data dimensions.

We opted to use a global gamma analysis to quantify our model in addition to traditional plan quality metrics (e.g., tumor coverage, dose conformity) since the predicted 3D dose distribution is not only limited to targets. Furthermore, in GKRS, metrics like coverage and conformity break down especially for small targets, as there are only a few voxels, thus making the metrics sensitive to small perturbations. Since large dose fall off is common in GKRS plans, global gamma was chosen instead of local gamma as it is less likely to exaggerate the errors in regions with high gradient [24]. As seen in the sub-analysis, our model performs best at predicting dose to voxels outside of the target area and worst on the periphery of the target as one would expect given the sharpness of the gradients there. While the predictions within the tumor were only marginally better than the periphery, the variation of dose within the tumor is usually not considered when evaluating treatment plans with the traditional plan quality metrics [25]. On the other hand, the result of the sub-analysis indicates that additional tuning of the models should be done to improve the predicted periphery dose, which would likely lead to an improvement to the coverage, specificity, and conformity of the predicted doses.

This approach has three notable limitations. First, we used a heterogenous dataset comprised of clinical plans that had a range in target sizes, prescription doses, number of isocenters, and number of targets (see **Figure 2**). For example, 3.7% of the tumor spaces in the dataset contained more than one target. As a result, the model may be less effective for patients with uncommon characteristics (e.g., patients with multiple nearby targets). Second, organs-at-risk were not considered in the models. Including organ-at-risk contours in the future would likely improve the prediction quality by directing more attention of the model towards important healthy tissue. Finally, all our training and testing data was modified via spline interpolation, which makes the model quality dependent on the size of interpolation errors. As a result, poorly interpolated data could have adverse effects that limit the model performance in both the training and testing processes.

## 5. Conclusion

In this study, we developed a novel KBP method for GKRS, supported by a data modification pipeline that transforms and upscales GKRS patient data for usage in machine learning-based 3D dose prediction. We demonstrate that utilizing the augmented data enables standard neural network models to produce high quality dose predictions for GKRS patients that are superior to existing state-of-the-art techniques. The resulting predictions have the potential to support the development of high-quality treatment plans as part of an automated KBP pipeline.

## 6. Acknowledgements

This research did not receive any specific grant from funding agencies in the public, commercial, or not-for-profit sectors.